\documentclass[10pt,conference]{IEEEtran}
\usepackage{color}
\usepackage{verbatim}
\usepackage{afterpage}
\usepackage{url}
\usepackage{xfrac}
\usepackage{algorithm}
\usepackage{algorithmicx}
\usepackage{algpseudocode}
\usepackage{amsmath}
\usepackage{amsfonts}
\usepackage{amssymb}
\usepackage{braket}
\usepackage{graphicx}
\usepackage{braket}
\usepackage[nocompress]{cite}
\usepackage{booktabs}
\usepackage{multirow}
\usepackage{caption}
\usepackage{subcaption}
\usepackage{cite}
\usepackage{diagbox}

\usepackage[utf8]{inputenc}
\usepackage[english]{babel}
\usepackage{abstract}

\usepackage{algpseudocode}
\usepackage{setspace}

\long\def\comment#1{}

\def\parah#1{\vspace*{0.0in} \noindent{\bf #1:}}

\IEEEoverridecommandlockouts
\DeclareMathOperator*{\argmin}{argmin}
\DeclareMathOperator*{\argmax}{argmax}

\newcommand{\re}{\mathrm{Re}}
\newcommand{\img}{\mathrm{Img}}
\newcommand{\tr}{\mathrm{Tr}}
\newcommand{\cE}{\mathcal{E}}
\newcommand{\qfacs}{$QFactor\ $}
\newcommand{\qfac}{$QFactor$}

\begin{document}
\title{\qfac: A Domain-Specific Optimizer for Quantum Circuit Instantiation}

\author{
\IEEEauthorblockN{Alon Kukliansky\IEEEauthorrefmark{1}, Ed Younis \IEEEauthorrefmark{4}, Lukasz Cincio\IEEEauthorrefmark{3}, Costin Iancu\IEEEauthorrefmark{4} }
\IEEEauthorblockA{
\IEEEauthorrefmark{1}
\textit{Naval Postgraduate School} 
\texttt{alon.kukliansky.is@nps.edu}}

\IEEEauthorblockA{
 \IEEEauthorrefmark{4}
\textit{Lawrence Berkeley National Laboratory} 
\texttt{\{edyounis,cciancu\}@lbl.gov}
}


\IEEEauthorblockA{
  \IEEEauthorrefmark{3}
\textit{Los Alamos National Laboratory} 
\texttt{lcincio@lanl.gov}
}

}

\newcommand{\lc}[1]{{\color{blue}{{#1}}}}

\maketitle

\begin{abstract}

We introduce a domain-specific algorithm 
for numerical optimization operations used by quantum circuit
instantiation, synthesis, and compilation methods. \qfacs uses a
tensor network formulation together with analytic methods and an
iterative local optimization algorithm to reduce the number of problem
parameters. Besides tailoring the optimization process, the
formulation is amenable to portable parallelization across CPU and
GPU architectures, which is usually challenging in general purpose
optimizers (GPO). Compared with several GPOs, our algorithm
achieves exponential memory and performance savings with similar
optimization success rates. While GPOs can handle directly circuits of
up to six qubits, \qfacs can process circuits with more than 12
qubits. Within the BQSKit optimization framework, we enable 
optimizations of 100+ qubit circuits using gate deletion algorithms
to scale out linearly with the hardware resources allocated for compilation in
GPU environments.
  
\end{abstract}

\section{Introduction}
\label{sec:introduction}

Quantum compilers have many purposes, but none are more critical than
reducing circuit gate count. This goal is especially true in the Noisy
Intermediate-Scale Quantum (NISQ) era, where each gate can have a high
cost in additional error rates. Compilation approaches that use
numerical optimization, commonly named {\it instantiation}-based
methods, are heavily used in quantum program development and
optimization. Hybrid quantum-classical algorithmic workflows such as
VQE~\cite{vqe} and QAOA~\cite{qaoa} repeatedly instantiate
parameterized quantum circuit (PQC) ansatz, which have gates
represented in their parameterized form, e.g., $R_x(\theta)$. Circuit
instantiation is directly supported by popular compilation
infrastructures such as IBM Qiskit~\cite{qiskit}, Google
Cirq~\cite{cirq}, or Tket~\cite{tket}. Instantiation is also an
important step in circuit synthesis tools such as
BQSKit~\cite{bqskit,qfast,qsearch,qce22_ed}, NACL~\cite{nacl},
Squander~\cite{squander1, squander2}, and CPFlow~\cite{cpflow}.

Quantum compilation and synthesis algorithms that utilize numerical
optimizers to perform instantiation follow a simple paradigm. First,
they will propose a candidate parameterized circuit template and,
second, use numerical instantiation to "best" fit the circuit
parameters to the target. One can find a solution by repeating this
process as often as necessary. The scalability of the underlying
general-purpose numerical optimizer (GPO) limits the scalability of
this compilation strategy. This limitation, in turn, limits the number
of qubits and gates that instantiation-based methods can directly
handle.

This paper makes the following contributions:
\begin{itemize}
\item We introduce Quantum Fast Circuit Optimizer (\qfac), a domain-specific optimizer for quantum circuit instantiation, usable in multiple custom or generic compilation workflows, designed to improve their scalability and efficacy.
\item We enable optimization approaches based on gate deletion heuristics to scale out when running in hybrid (CPU+GPU) distributed memory environments. 
\item  We provide valuable feedback about the strategy to architect instantiation-based compilation workflows able to handle huge circuits.
\end{itemize}

The main idea behind \qfacs is based on lessons learned from machine
learning algorithms that heavily use tensor network formulations,
combined with our experience developing scalable quantum synthesis
infrastructures. Tensor network algorithms have been also extensively used to simulate quantum circuits~\cite{guo2021verifying, nguyen2022tensor, pan2022simulation, chen2022quantum, vincent2022jet}. All existing instantiation-based synthesis approaches~\cite{bqskit,qfast,qsearch,qce22_ed,nacl,squander1,squander2,cpflow,PhysRevResearch.5.023146} use general-purpose optimizers and  parameterized gate representations. For example, the parameterization for a rotation is:

\begin{equation}
R_Y(\theta)=e^{-i \theta Y/2}=
\begin{pmatrix}
\cos{\frac{\theta}{2}} & -\sin{\frac{\theta}{2}} \\
\sin{\frac{\theta}{2}} & \cos{\frac{\theta}{2}}
\end{pmatrix}
\end{equation}

In contrast, \qfacs uses a tensor network formulation, which does not
require explicit parameterization, enabling the algorithm to work at
the unitary rather than the parameter level. This perspective change
drastically reduces the total number of optimized parameters compared
to GPOs, because each gate may have many parameters but only one
unitary.

We provide a CPU-based implementation written in Rust, together with a
Python implementation written using JAX~\cite{jax2018github}. The Rust
implementation is serial, while the Python/JAX implementation benefits
from auto-parallelization. For validation we use a suite of algorithms whose implementation ranges from five to 400 qubits, with a total gate count as high as $\approx 170,000$ gates, as shown in
Table~\ref{tab:benchmarks}.

We first evaluate the instantiation performance of \qfacs against
Ceres~\cite{Agarwal_Ceres_Solver_2022} and
LBFGS~\cite{nlopt,liu1989limited,nocedal1980updating}.
 The GPO's have good performance for
circuits up to 6-7 qubits, while \qfacs can process circuits with more
than 12 qubits. Best performance is achieved on GPU based systems,
which outperform CPUs by 4$\times$ in the bigger circuits. Scalability
is problem dependent, and we conjecture that the new formulation is
limited only by GPU memory size. The success rate of \qfacs is similar
to that of GPOs for circuits evaluated.

To leverage \qfac's advantages in circuit instantiation, we
incorporated it into BQSKit`s~\cite{bqskit} gate deletion based
optimization~\cite{qce22_ed} pipeline, which can handle circuits
with 100s of qubits. To ensure scalability for large qubit and depth
count circuits, BQSKit partitions these into smaller panels
which are then directly optimized.  When comparing configurations of this
approach using GPOs or \qfac, \qfacs improves the optimized circuit
quality by enabling the usage of bigger partitions, thus reducing U3
and CNOT gate count, in some cases, by more than $30\%$ . Furthermore,
\qfacs appears to scale linearly with resources on GPU based systems,
while all the other CPU based implementations suffer from load
balancing problems.

Given their efficacy in optimizing circuits, infrastructures similar
to BQSKit are very attractive for NISQ systems, where gate count and
circuit depth affect fidelity.  The capability to directly instantiate
bigger circuits enables us to draw very useful conclusions to guide
the architecture of such hierarchical transformation  approaches that combine
partitioning and instantiation. The memory footprint of instantiation
scales exponentially with qubits, and it is therefore limited by the
memory capacity of a system. At the high-end qubit count (e.g. IBM
Osprey with 433 qubits) a legitimate question is whether we should
strive to develop even more scalable direct instantiation methods or have we
already reached diminishing returns. Our experimental data empirically  indicates
that the latter might be the case: benefits seem to saturate when using partitions with more than 8-9.

Besides the benefits showcased in this paper, \qfacs is definitely
very useful and enables other optimization approaches. We have
formulated the prototype algorithm in Python circa 2021 and have since used it in optimization of convolutional neural networks and unitary compilation~\cite{caro2022generalization}, state preparation~\cite{eckstein2023large}, learning fast scramblers~\cite{caro2022out} and in conjunction with variable ansatz techniques~\cite{bilkis2021semi}. Furthermore, as the
performance gap between GPUs and CPUs is likely to increase, the
availability of a scalable domain specific optimizer for instantiation
bodes well for the future of hierarchical synthesis approaches.

The organization of this paper is as follows. Section~\ref{sec:background} introduces the background concepts. Section~\ref{sec:alg} provides a detailed description of the algorithm, while Section~\ref{sec:implementation} covers all the implementation details. The evaluation methods and results are presented in Section~\ref{sec:evaluation}, and the discussion of the results is presented in Section~\ref{sec:discussion}. Finally, Section~\ref{sec:conclusion} provides the conclusion of the paper.

\section{Background}
\label{sec:background}

\begin{figure}
    \centering
    \includegraphics[scale=1.5]{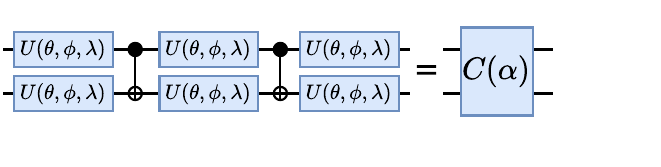}
    \caption{\footnotesize \it This is an example of a parameterized
      quantum circuit on the left. It is composed of three-parameter
      universal single-qubit rotations and two-qubit CNOT gates. For
      direct simulation, we can represent the circuit by its
      unitary operator shown on the right, which is calculated by
      tensor contraction of all of its gates. Furthermore, we can
      represent parameterized circuits by a parameterized unitary
      $C(\alpha)$, which can be instantiated to some other unitary $V$
      by solving for the parameters $\alpha$ that maximize the overlap
      of $C(\alpha)$ and $V$. This can be accomplished with analytic
      methods in specific cases and gradient descent or other
      numerical methods in the general case. }
    \label{fig:params}
\end{figure}

\parah{Parameterized Circuit Instantiation} Instantiation is the process of finding the parameters for
a circuit's gates that make it to most closely implement
a target unitary.
For a $n$-qubit parameterized quantum circuit $C: \mathbb{R}^k \mapsto U(N)$ and a target unitary $V
\in U(N)$, where $N=2^n$, solve for $$\argmax_{\alpha}{|\mathrm{Tr}(V^{\dagger}C(\alpha))|}$$
where $k$ is the number of gate parameters in the circuit, and $U(N)$
is the set of all $N \times N$ unitary matrices. This definition is
very general and considers the parameterized circuit as a
parameterized unitary operator, see Figure~\ref{fig:params}. The
$|\mathrm{Tr}(V^\dagger C(\alpha))|$ component measures the Hilbert-Schmidt inner
product, which physically represents the overlap between the target
unitary and the circuit's operator. The maximum value this can have is
equal to $N$, the dimension of the matrix, and this occurs when
$C(\alpha)$, the unitary of the circuit with gate parameters $\alpha$,
is equivalent to $V$ the target unitary up to a global phase.

 Techniques that perform instantiation are ubiquitously
deployed in quantum compiler toolchains from industry and academia. The most common form  is the KAK~\cite{kak}
decomposition, which uses analytic methods to produce the two-qubit
circuit that implements any two-qubit unitary. KAK uses a native gate set composed of single-qubit rotations and CNOT gates.

Recently, bottom-up approaches to quantum synthesis have been
successful through numerical instantiation~\cite{qsearch, qfast, nacl,
  squander1, squander2, bestapprox}. When compared to KAK, they handle
circuits of more than two qubits, roughly up to 6-7 qubits, albeit
with large runtime overhead.  When compared against commercial
compilers such as Qiskit, Cirq,  and Tket, direct synthesis
techniques have been shown to provide  higher quality
circuits. Rather than fixed mathematical identities, these techniques
employ a numerical optimizer to closely approximate a solution to the
instantiation problem. This is done by minimizing a cost function,
often the unitary error or distance between the circuit's unitary and
a target unitary. This is given by the following formula using the
same notation as before.

$$\Delta(C(\alpha), V) = 1 - \frac{|\mathrm{Tr}(V^\dagger C(\alpha))|}{N}$$

Other variations of this distance function include:

$$\Delta_f(C(\alpha), V) = 1 - \frac{\mathrm{Re}(\mathrm{Tr}(V^\dagger C(\alpha)))}{N}$$

and

$$\Delta_p(C(\alpha), V) = \sqrt{1 - \frac{|\mathrm{Tr}(V^\dagger C(\alpha))|^2}{N^2}}$$

All three cost functions have a range of $[0, 1]$, and as they approach zero,
the circuit's unitary approaches $V$.

\parah{Numerical Optimization in Instantiation} Synthesis tools can
use either derivative-free or gradient-based general purpose
optimizers when the problem formulation allows it. For example, Davis
et al~\cite{qsearch} discuss derivative-free (CMA-ES, COBYLA, and
BOBYQA) and gradient-based optimization (BFGS and Levenberg-Marquardt)
for a synthesis algorithm that uses $U3(\theta,\phi,\lambda)$
parameterized single-qubit gates.  They report that the
Ces~\cite{Agarwal_Ceres_Solver_2022} implementation of Levenberg-Marquardt with
gradients provides up to $100\times$ execution time improvements when
compared to the implementation of COBYLA provided in {\tt scipy}, with
better scaling as the number of variables increases.  Many quantum
compiling frameworks utilize gradient-descent optimizers such as
L-BFGS~\cite{liu1989limited,nocedal1980updating} and least-squares optimizers such as Levenberg
Marquardt~\cite{ranganathan2004levenberg} for certain compilation objectives.

Lavrijsen et al~\cite{vqeopt} provide a thorough study of optimizers
appropriate for hybrid quantum-classical variational algorithms, which
perform instantiation in their classical step. They evaluate (hybrid)
mesh algorithms (ImFil~\cite{imfil}, NOMAD~\cite{nomad}); local fit
(SnobFit~\cite{snobfit}); and trust region algorithms
(PyBobyqua~\cite{boby1,boby2}).

Overall, all these instantiation studies agree that the scalability of
the numerical optimizer with circuit size is the biggest bottleneck. Improving it can
only lead to shorter time to solution with better quality results. Based on
our previous experiences and literature recommendations, in this study
we choose to compare against the gradient-descent L-BFGS~\cite{nocedal1980updating,liu1989limited}
optimizer and the Levenberg-Marquardt~\cite{ranganathan2004levenberg} least-squares
optimizer.

\begin{figure}[htbp!]
    \includegraphics[width=\columnwidth]{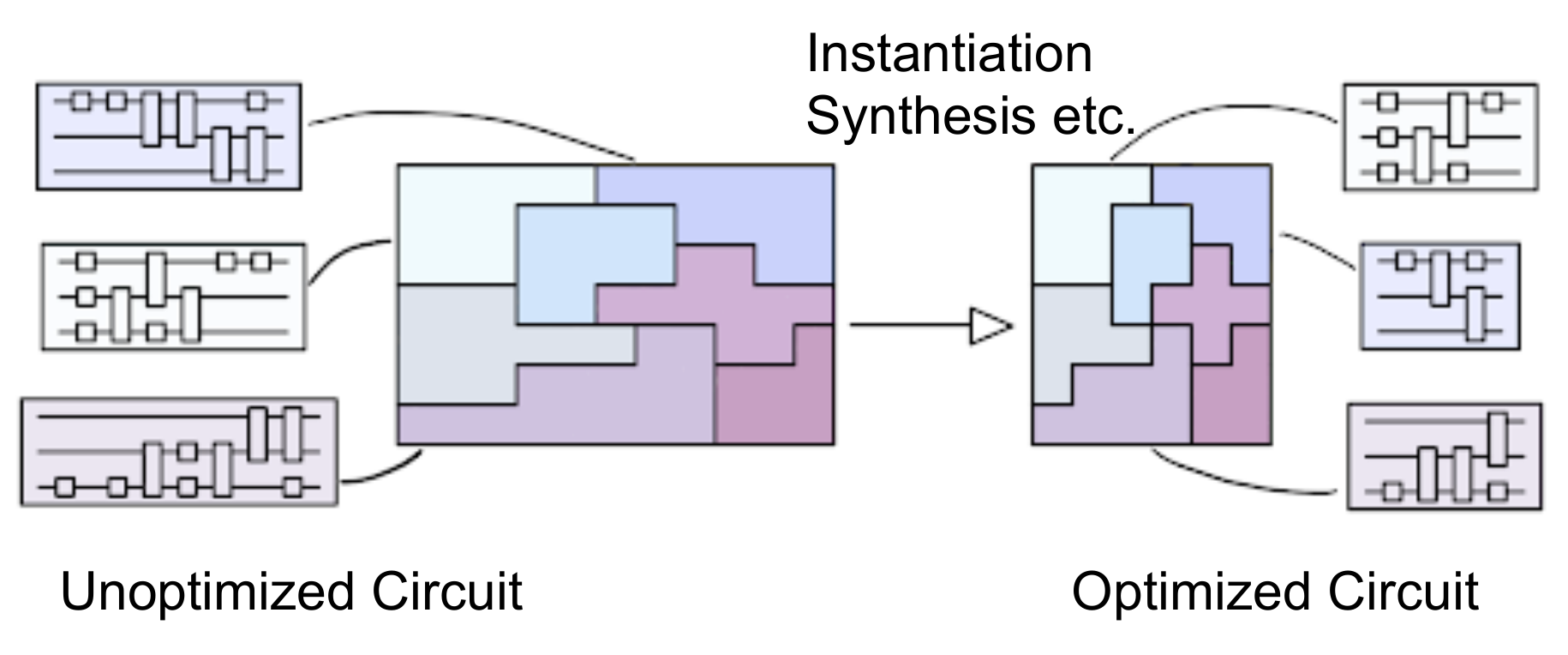}
    \caption{\footnotesize \it Hierarchical synthesis or instantiation based circuit optimization. Both partitioning strategy and quality of numerical optimization determine
    performance and final output quality.}
    \label{fig:partf}
\end{figure}

\parah{Partitioning and Resynthesis}
Approaches that numerically optimize the parameters of a circuit by
simulating it encounter exponential-scaling issues with the increase in qubits and circuit depth. As a result,
direct methods are often limited to low dimensionality problems with few qubits and shallow circuits.
Hierarchical approaches can overcome these scaling
issues. These methods will group closely located gates into
fixed-width blocks and then resynthesize or transform them with numerical
optimization-based strategies. This process is illustrated in Figure~\ref{fig:partf}.

Approaches that follow this similar partitioning-driven flow, such as
QGO~\cite{qgo} and BQSKit~\cite{bqskit}, have been shown to scale to
1000s of qubits, but could only use three qubit partitions due to slow
numerical optimization.  Their behavior is likely influenced by the
block size for both result quality and completion time. The number of
qubits chosen per block makes an impact, because more qubits are
likely to capture more domain-level physical interactions, leading to
better optimization opportunities.  On the other hand, larger blocks require more memory
since they need to be represented by exponentially larger matrices,
and they require more processing time due to the increase in the number of
parameters. Improving
instantiation is likely to lead to improved compilation performance,
as well as output circuit quality.

\section{Algorithm} \label{sec:alg}
 \qfacs simplifies the circuit's parameter complexity by
directly updating (possibly multi-qubit) unitary gates without
parameterizing them internally. The optimization is done by performing an
iterative sweep through all the gates. During every step, \qfacs
locally optimizes each gate. This ability to treat each gate as a
unitary without parameterization during optimization sets \qfacs apart
from other general-purpose optimizers and makes it particularly
effective.

 Given
$2^n \times 2^n$ unitary $V$, \qfacs finds unitaries $u_1, \ldots, u_p$ such
that the Froebenius norm  between $V$ and unitary $U \equiv
u_1 \cdot \ldots \cdot u_p$ is minimized. 

\begin{equation} \label{eq:cost}
|| U - V ||^2 = 2^{n+1} \left( 1 - \frac{1}{2^n} 
\re \tr (V^\dagger U)
\right) \ .
\end{equation}

\begin{figure}[t]
    \includegraphics[width=\columnwidth]{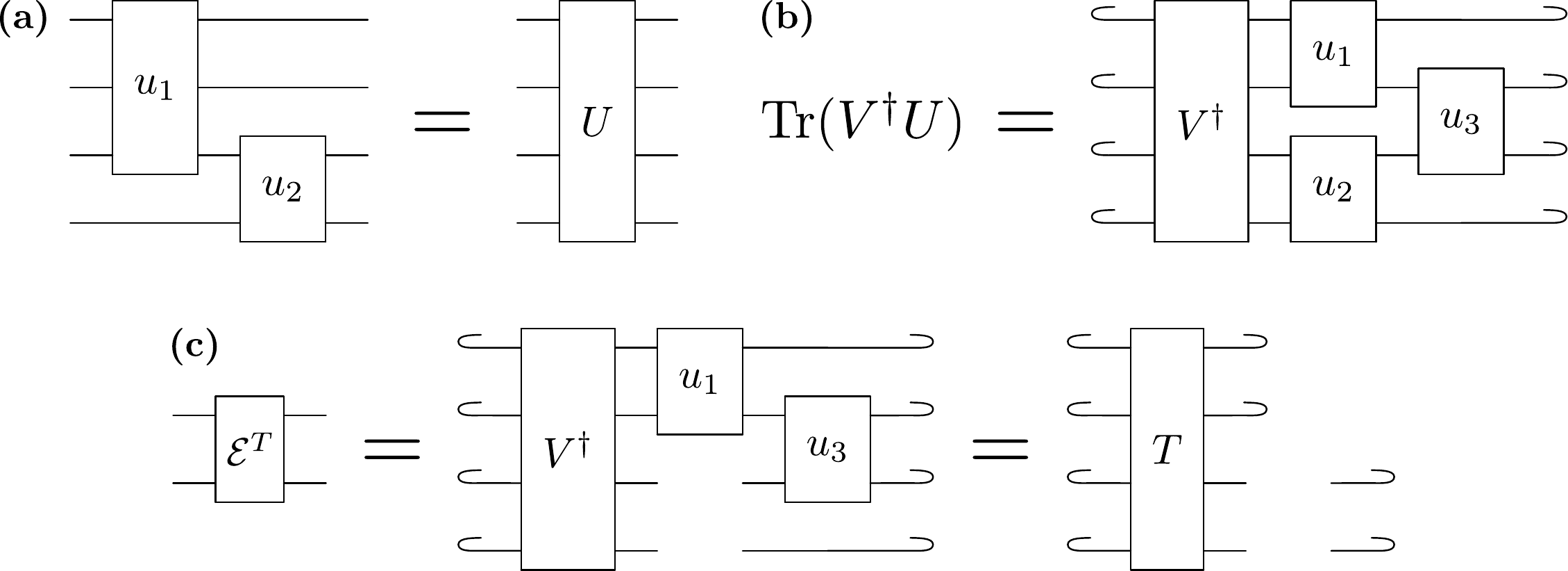} \caption{\it \footnotesize Tensor
    network calculations used in \qfac. (a)~An example of a tensor
    contraction. Three-qubit gate $u_1$ is contracted with two-qubit
    gate $u_2$ to form gate $U = u_1 \cdot u_2$. (b)~Four-qubit
    example of a cost function in Eq.~\eqref{eq:cost} (up to constant
    factors). Here, $U = u_1 \cdot u_2 \cdot u_3$. The value of
    $\mathrm{Tr}(V^\dagger U)$ is represented as a tensor
    network. Bent qubit line represents partial
    trace. $\mathrm{Tr}(V^\dagger U)$ can be written as
    $\mathrm{Tr}(\mathcal{E} u_2)$, where the transpose of environment
    matrix $\mathcal{E}$ is defined in~(c). Tensor $T$ is constructed
    in the process of computing environment matrix $\mathcal{E}$, $T =
    u_3 \cdot V^\dagger \cdot u_1$.  } \label{fig:tensors}
\end{figure}

Note that the ``$\cdot$'' in $u_1 \cdot u_2$ represents tensor
contraction, as shown in Fig.~\ref{fig:tensors}(a). Formally, the
operation is properly defined only when information on all qubits participating in any
$u_k$  is given. We drop that information for simplicity. Since all
gates $u_k$'s enter Eq.~\eqref{eq:cost} linearly, $\mathrm{Tr}
(V^\dagger U)$ can be written as $\mathrm{Tr} (\mathcal{E} u_k)$,
where $\mathcal{E}$, so-called environment
matrix~\cite{orus2014practical}, does not depend on $u_k$, see
Fig.~\ref{fig:tensors}(b) and (c). This fact is used in optimization
over $u_k$'s, as described later in Section~\ref{sec:optimization}.

The problem is thus to find
\begin{equation}
\argmin_{ \{u_k\} } || u_1\cdot\ldots\cdot u_p -V ||
\end{equation} 

or equivalently
\begin{equation} \label{eq:Vu}
\argmax_{ \{u_k\} } \re\tr (V^\dagger u_1\cdot\ldots\cdot u_p) \ .
\end{equation}

\subsection{Optimization Algorithm} \label{sec:optimization}

The algorithm uses the fact that the local optimization problem
\begin{equation} \label{eq:argmax}
\argmax_{u \mathrm{- unitary}} \re \tr ( \cE u )
\end{equation}
can be solved exactly for any arbitrary environment matrix~$\cE$. As shown in 
Alg.~\ref{alg:qfactor}, \qfacs iteratively calculates
the environment matrix for each gate and optimizes its unitary, thus
gradually improving the implementation of the circuit.

\qfacs updates unitaries $u_k$ sequentially and individually. It
starts with $u_1$ and proceeds to the right, updating $u_p$ in the
final step. It then updates the unitaries in the opposite
direction. This sequence of updates forms one iteration of the
algorithm and is called TwoSidedSweep. Every individual update of a
given $u_k$ is done by calculating it's enviorment, see
Fig~\ref{fig:tensors}(c) and then using the SVD update described and
proved in Section~\ref{sec:local_opt_proof}. This update is referenced
as OptimizeGate in Alg.~\ref{alg:qfactor}.

Naively, one needs to recalculate each environment matrix for each of
the gates after a gate update. \qfacs has an optimization that
reduces both the memory and runtime complexity of these calculations
by reusing previous calculations. It represents $V^\dagger U$ as a
tensor, which we call the circuit tensor. \qfacs sweeps over the
gates with a ``peephole'' by applying the gates or their inverse to
the left or to the right of the tensor according to the algorithm
phase. Applying a gate is simply contracting the circuit tensor with
the gate tensor, as seen in Fig.~\ref{fig:tensors}(a). Applying the
inverse of a gate removes it from the tensor, and by tracing all the
other legs of the circuit tensor we create the environment or
''peephole'', as can be seen in Fig~\ref{fig:tensors}(c). The initial
creation of the circuit tensor and the sweeping procedure can be seen
in Alg.~\ref{alg:qfactor} at InitCircuitTensor and TwoSidedSweep
respectively. The tracing procedure is referenced as CalcEnvMat in
Alg.~\ref{alg:qfactor}.

The \qfacs algorithm is designed such that every iteration does not
increase the cost function; rather, it can only decrease or reach a
plateau. To facilitate termination, the algorithm employs a mechanism
composed of multiple criteria. The primary criterion for termination
is achieving the goal of instantiating the parameterized circuits to
the target unitary within a given tolerance. Additionally, the
algorithm terminates when it detects that it has reached a plateau or
when it has completed a maximum number of iterations. For a detailed
explanation of the hyperparameters that influence the algorithm
behavior, please refer to Section~\ref{sec:hyper_params}.

\subsubsection{Local Optimality Proof}\label{sec:local_opt_proof}

Let us write singular value decomposition of $\cE$ as $\cE =
XDY^\dagger$, with unitary $X$, $Y$ and diagonal, real, non-negative
$D$.

We have
\begin{align}
\re \tr ( \cE u ) &= \re \tr (X D Y^\dagger u) \nonumber \\
&=\re \tr (D Y^\dagger u X) = \re \tr (DW) 
\end{align}
for some unitary $W$.

Finally,
\begin{equation}
\re \tr (DW) = \sum_j D_{jj} \re W_{jj} \leq \sum_j D_{jj} \ .
\end{equation}

$\sum_j D_{jj} \re W_{jj}$ is maximized by setting all $W_{jj} = 1$. $W$ is unitary, so the maximum is achieved at $W = \mathrm{I}$, the identity matrix. This means that the unitary that solves \eqref{eq:argmax} is given by
\begin{equation} \label{eq:opt_u}
    u_{\textrm{new}} = YX^\dagger \ .
\end{equation}

\subsubsection{\qfac's Hyperparameters}\label{sec:hyper_params}

\qfacs has the following hyperparameters that control the termination
conditions of the algorithm, and update policy and randomization:
\begin{itemize}
    
    \item \textit{dist\_tol} - When the distance between the target unitary to the current circuit reaches \textit{dist\_tol}, the algorithm stops. We calculate the distance as $|| u_1 \cdot \ldots \cdot u_k - V ||$.
    
    \item \textit{diff\_tol\_a} and \textit{diff\_tol\_r} - The algorithm will terminate if $|c_{i}| - |c_{i-1}| \le \textit{diff\_tol\_a} + \textit{diff\_tol\_r} * |c_{i}|$ where $c_{j}$ is the cost function value after iteration $j$.
    
    \item \textit{long\_diff\_count} and \textit{long\_diff\_r} - Control the long plateau detection mechanism.
     The algorithm will terminate if over the course of \textit{long\_diff\_r} iterations, the cost function did not decrease by \textit{long\_diff\_r}, i.e. the condition
    \begin{equation*}
        |c_{i}| - |c_{i-\textit{long\_diff\_count}}| \le \textit{long\_diff\_r}  * |c_{i-\textit{long\_diff\_count}}| 
    \end{equation*}
    must be met to continue the optimization. Here $c_{j}$ is the cost function value after iteration $j$.

  \item \textit{min\_iter} - Sets the minimum amount of iterations for \qfacs to complete before stopping.
    
    \item \textit{max\_iter} - Sets the maximum amount of
    iterations. The algorithm will always stop when it reaches this
    limit.

   \item \textit{reset\_iter} -  In order to conserve memory, the algorithm (effectively) performs
operations that should evaluate to identity. For example, it applies
gate $u$ to compute environment and then applies $u^\dagger$ at a
later step. This builds rounding error over many iterations, which can
be observed in practice.  Thus, for numerical stability, we reset the
circuit tensor every \textit{reset\_iter} iterations. Choosing small values for this
configuration parameter has a performance impact. From our experience
setting it to  40, has minimal performance impact and provides
a  wide safety margin.

    \item \textit{multistarts} and \textit{seed} - To overcome the local minimum problem, one can run \qfacs with various initial gate unitaries, in the hope that at least one of the runs will lead to a good solution. The initial random unitaries for each gate are controlled by a \textit{seed} parameter.

    \item \textit{Beta} - Is a regularization parameter that controls
    the retention of the old value in the gate update step. Instead of
    performing the SVD operation on the environment $\cE$, it is done
    on \[ (1-\beta)\cE + \beta * u^\dagger \] This parameter is useful
    in overcoming slow convergence for circuits that have
    dependency between gates, and the local optimization fails to
    realize it. At the limits, setting $\beta = 0$ corresponds to a
    ``full'' update described above, while $\beta = 1$ results in no
    update.
    
\end{itemize}

From our experience, here are some good initial hyperparameters values: \textit{dist\_tol} $=10^{-10}$, \textit{diff\_tol\_a} $=0$, \textit{diff\_tol\_r} $=10^{-5}$, \textit{long\_diff\_count} $=100$, \textit{long\_diff\_r} $=0.1$, \textit{min\_iter} $=0$, \textit{max\_iter} $=10^{5}$, \textit{reset\_iter} $=40$, \textit{multistarts} $=8$, $\beta=0$.
By modifying the above mentioned hyperparameters, one can easily adjust the tradeoff between result quality and execution time. 
If one wishes to get results faster, decrease \textit{max\_iter}, increase
\textit{diff\_tol\_r} and \textit{diff\_tol\_a}. Alternately, one might increase $\beta$ or \textit{multistarts} to find
better results with additional execution overhead.
    
\subsubsection{Specializing Gates Sets}
Besides using general unitaries as the gate set, \qfacs can use more  specific parameterized gates. 
For example, one can target a trapped ion quantum computer that
has the single qubit gates
$GPI(\theta)$ \cite{wrightBenchmarking11qubitQuantum2019c} and
$R_z(\theta)$ and the Mølmer-Sørenson entangling gate
$MS(\phi_1, \phi_2)$ \cite{MS.PhysRevLett.82.1835}.  In those
situations, the SVD approach isn't viable and we need to directly find
$\argmax_\theta \re \tr ( \cE G(\theta ))$. This is an easy task, as
we can analytically calculate a formula that gives the maximum for a
given environment.  For example, let's examine the calculation needed
for optimizing the $R_z$ gate. The unitary of $R_z(\theta)$ is given
by
\begin{equation*}
Rz(\theta) = \left(
        \begin{array}{cc}
            1 & 0 \\
            0 & e^{i\theta}
        \end{array} \right)
\end{equation*}
Hence, we need to find $\theta$ that maximizes
\begin{align*}
    \re \tr ( \cE Rz(\theta ))  &= \re \tr \left(\left(
    \begin{array}{cc}
            \cE_{0,0} & \cE_{0,1} \\
            \cE_{1,0} & \cE_{1,1}
        \end{array} \right) *  
        \left( \begin{array}{cc}
            1 & 0 \\
            0 & e^{i\theta}
        \end{array} \right) \right) \\
        &= -\re(\cE_{1,1})\cos{\theta} + \img(\cE_{1,1})\sin{\theta}
\end{align*}
When $\re(\cE_{1,1})>0$ and $\img(\cE_{1,1})\ne0$ the maximum is achieved by:
\begin{equation*}
    \theta_{\textrm{new}} = \arctan{\frac{\re(\cE_{1,1})}{\img(\cE_{1,1})}}
\end{equation*}

\begin{algorithm}
\caption{\qfac}
\label{alg:qfactor}
\begin{algorithmic}

\Function{InitCircuitTensor}{target unitary, unitaries, locations}
\State $ct \gets reshaped~target~unitary$

\For{$(u, l) \in (unitaries, locations)$}
    \State $ct \gets \Call{ApplyRight}{ct, u, l, inverse=False}$
\EndFor

\State \textbf{return} $ct$
\EndFunction

\vspace{0.5\baselineskip} 

\Function{TwoSidedSweep}{circuit tensor, unitaries, locations, gates}
\State $ct \gets \text{circuit tensor}$
\State $newUs$ $\gets$ empty list
\For{$(u, l, g) \in reversed(unitaries, locations, gates)$}
    \State $ct \gets \Call{ApplyRight}{ct, u, l, inverse=True}$
    \State $env \gets \Call{CalcEnvMat}{ct, l}$
    \State $u_{opt} \gets \Call{OptimizeGate}{g, env} $
    \State $newUs$ $\gets$ \Call{PrependtoList}{$newUs$, $u_{opt}$}
    \State $ct \gets \Call{ApplyLeft}{ct, u_{opt}, l, inverse=False}$
\EndFor

\vspace{0.2\baselineskip}

\State $finalUs$ $\gets$ empty list
\For{$(u, l, g) \in (newUs, locations, gates)$}
    \State $ct \gets \Call{ApplyLeft}{ct, u, l, inverse=True}$
    \State $env \gets \Call{CalcEnvMat}{ct, l}$
    \State $u_{opt} \gets \Call{OptimizeGate}{g, env} $
    \State $finalUs$ $\gets$ \Call{AppendtoList}{$finalUs$, $u_{opt}$}
    \State $ct \gets \Call{ApplyRight}{ct, u_{opt}, l, inverse=False}$
\EndFor

\vspace{0.2\baselineskip}
\State $\textbf{return}$ $ct$, $finalUs$

\EndFunction

\vspace{0.5\baselineskip} 

\Function{Qfactor}{target unitary, initial unitaries, locations, gates, \textit{termination conditions}}
\State $ct \gets$ \Call{InitCircuitTensor}{target unitary, initial unitaries, locations}
\State $unitaries \gets$ initial~unitaries
\Repeat 
    \State $ct$, $unitaries \gets$  \Call{TwoSidedSweep}{$ct$, $unitaries$, locations, gates}
    \State calculate cost function and detect a plateau 
\Until{\textit{termination conditions} not met}

\vspace{0.2\baselineskip}
\State \textbf{return} $unitaries$
\EndFunction

\end{algorithmic}
\end{algorithm}

\section{\qfacs Implementation}
\label{sec:implementation}

 We implemented  \qfacs as a standalone optimizer in Python ({\tt
   \footnotesize https://github.com/BQSKit/qfactor}) and Rust ({\tt
   \footnotesize https://github.com/BQSKit/bqskitrs}).
These are serial CPU based implementations. We also
 provide a parallel version using JAX~\cite{jax2018github}. All
 \qfac's implementations have also  been incorporated into the BQSKit
 compilation infrastructure at {\tt \footnotesize
   https://github.com/BQSKit/bqskit}.

\subsection{Migrating to GPU using JAX}

GPUs have been shown to significantly outperform CPUs for tensor network formulations, due to their ability to execute more operations in parallel.
We have ported the \qfacs Python implementation to a GPU implementation using the JAX
framework~\cite{jax2018github}. This allows us to seamlessly parallelize
the instantiation across different multistarts and JIT our code to
better utilize the GPU, while still working at a high level of
representation in Python. To fully utilize the GPU, even when using
small partitions that don't saturate the GPU, we use NVIDIA's
MPS \cite{NVIDIA_MPS}, which enables different processes to share GPU
resources and dynamically allocate them according to the changing load
of each process.

In contrast to the CPU implementation, where the instantiation using
different starting conditions is done in sequence, on the GPU all the
multistarts are done in parallel - this raises the question of when
should the optimization be terminated. Of course, if one of the
multistarts was able to converge to the correct solution, we can stop
and return that result. The tricky case is termination by plateau
detection. As explained in Section~\ref{sec:alg}, \qfacs can get stuck
in local minima, so the heuristic of plateau detection needs to take
into account the parallel multistarts and terminate only if all of
them have hit a plateau.

\section{Evaluation}
\label{sec:evaluation}
We evaluate \qfacs on a set of benchmarks that represent real circuits
ranging from four  to 400 qubits and containing up to $\approx 200,000$ gates, as illustrated in
Table~\ref{tab:benchmarks}. We use circuits that implement classical operations (multiply, add)~\cite{hassan_qarithmetic_2021}, the HHL, Shor~\cite{shor1999polynomial} and Grover algorithms~\cite{grover}, variational algorithms (VQE and QAOA)~\cite{qaoa, hardware_efficient}, Quantum Phase Estimation circuits, Heisenberg and Hubbard models~\cite{bravyi_kitaev, hubbard}, and Transverse Field Ising Model circuits~\cite{tfimshin, bassman_arqtic_2021}.

After gathering all the benchmarks, we
standardized them to the U3 and CNOT gate-set. We accomplish that by
using the Qiskit~\cite{qiskit} compiler with no optimizations
(O0). The standardized benchmarks can be found in  the {\tt qce23} benchmark
repository on GitHub~\cite{qce23benchmarks}.

We compare against two leading general-purpose numerical optimizers,
which we will refer to as CERES~\cite{Agarwal_Ceres_Solver_2022} and
LBFGS~\cite{liu1989limited, nlopt, nocedal1980updating}. These are production quality, ubiquitously used
optimizers. Furthermore, their hyperparameters have been carefully tuned by Davis et al~\cite{qsearch,leap} for instantiation in BQSKit.

We study the performance of each implementation on both CPUs and GPUs,
and we benchmark serial execution, single device (CPU or GPU), as well as strong
scaling in distributed memory.  For strong scaling we use the BQSKit internal runtime support instead of relying
on JAX. Also, while JAX can support auto-parallelization for CPU, it's
performance is notoriously bad, also true for our experiments.
Therefore we do not evaluate \qfacs implemented in JAX on CPUs. Note also
that for large circuits that are partitioned, while the workload is
embarrassingly parallel, strong scaling behavior is ultimately
determined by load balancing.

We first compared the ability to solve instantiations for
each optimizer. In these experiments, we assign a time budget to each
instantiation request and measure the success rate (defined as finding
a solution) on a set of circuits. The results are presented in Section~\ref{sec:inst_eval}. 
To evaluate practical usage in circuit transformations, we also assess the quality of each optimizer based on the U3 and CNOT gate
reduction in circuits produced by a new BQSKit hierarchical gate deletion pass implemented using \qfac. This workflow is discussed in Section~\ref{sec:synth_eval}.

In our comparison, we denote the Rust and Python+JAX implementations
of \qfacs by QF-RUST and QF-JAX respectively. To distinguish
parallel CPU runs, we suffixed the instantiator name with '\_P'.

The evaluation shows that when comparing LBFGS against CERES, while
the former is slightly slower for serial execution on CPUs, they
exhibit similar quality for parallel executions. Therefore, we select
CERES for most comparisons, omitting detailed LBFGS results for
brevity.

\parah{Evaluation Setup}
Our evaluation was done on NERSC's Perlmutter\cite{perlmutter.arch}
supercomputer, which has two node flavors: hybrid GPU-CPU or pure
CPU. Each hybrid node has one AMD EPYC 7763 64-core processor,
256GB DDR4 DRAM, and four NVIDIA A100 40GB GPUs, while each CPU node
has two AMD EPYC 7763 64-core processors and 512GB DDR4 DRAM.  The
CPU-only workloads(CERES, LBFGS, and QFACTOR-RUST) were run on the CPU
nodes, while the GPU workload, JAX implementation of \qfacs, ran on
the GPU nodes. To better utilize the GPUs we setup an NVIDIA MPS
server for each GPU on the node, and each \qfacs process is
instructed to use only a single GPU, leveraging NVIDIA's
CUDA\_VISIBLE\_DEVICES environment variable.

Each instantiation operation is configured to sample {\it multistarts}=32 different starting points. For parallel execution on
CPUs we allocate all the cores available.  On GPUs we use the entire
device, but it requires a more careful decision about the degree of
parallelism allowed.  Here, there is a tradeoff between partition size
and the number of jobs allowed to execute in parallel. Partition
``volume'' (qubits and gate count) determines both the memory
footprint and the amount of hardware resources required for a single
job. For a given qubit count, shallow circuits are
likely\footnote{This depends on the code generation strategy. GPU JITs
are attempt to tile loops in this manner.}  to require fewer execution
units than deep circuits. After experimentation, for GPUs we have
chosen the set of defaults described in
Table~\ref{tab:workers_for_size}.

The other \qfacs hyperparameters used are: {\it dist\_tol = $10^{-10}$,
 diff\_tol\_a=0, diff\_tol\_r=$10^{-5}$,
long\_diff\_count=100, long\_diff\_r=$0.1$, min\_iter=0, reset\_iter=40, multistarts=32, seed=None,
Beta=0}.

\begin{table}[htbp!]
\footnotesize
\centering
\begin{tabular}{ |c|c| }
\hline
 Partition Size & Workers per GPU  \\ 
 \hline 
 3,4 & 10 \\
 5 & 8 \\
 6 & 4 \\
 7 & 2 \\
 8 and more & 1 \\ 
 \hline
\end{tabular}
\caption{\footnotesize \it Relation between the instantiated partition size and the amount of workers spawned for every GPU}
\label{tab:workers_for_size}
\end{table}

\subsection{Circuit Instantiation: Runtime and Success Rate}\label{sec:inst_eval}
\begin{figure*}[htbp!]
\begin{tabular}{cc}
\begin{minipage}{\columnwidth}
    \includegraphics[width=\columnwidth]{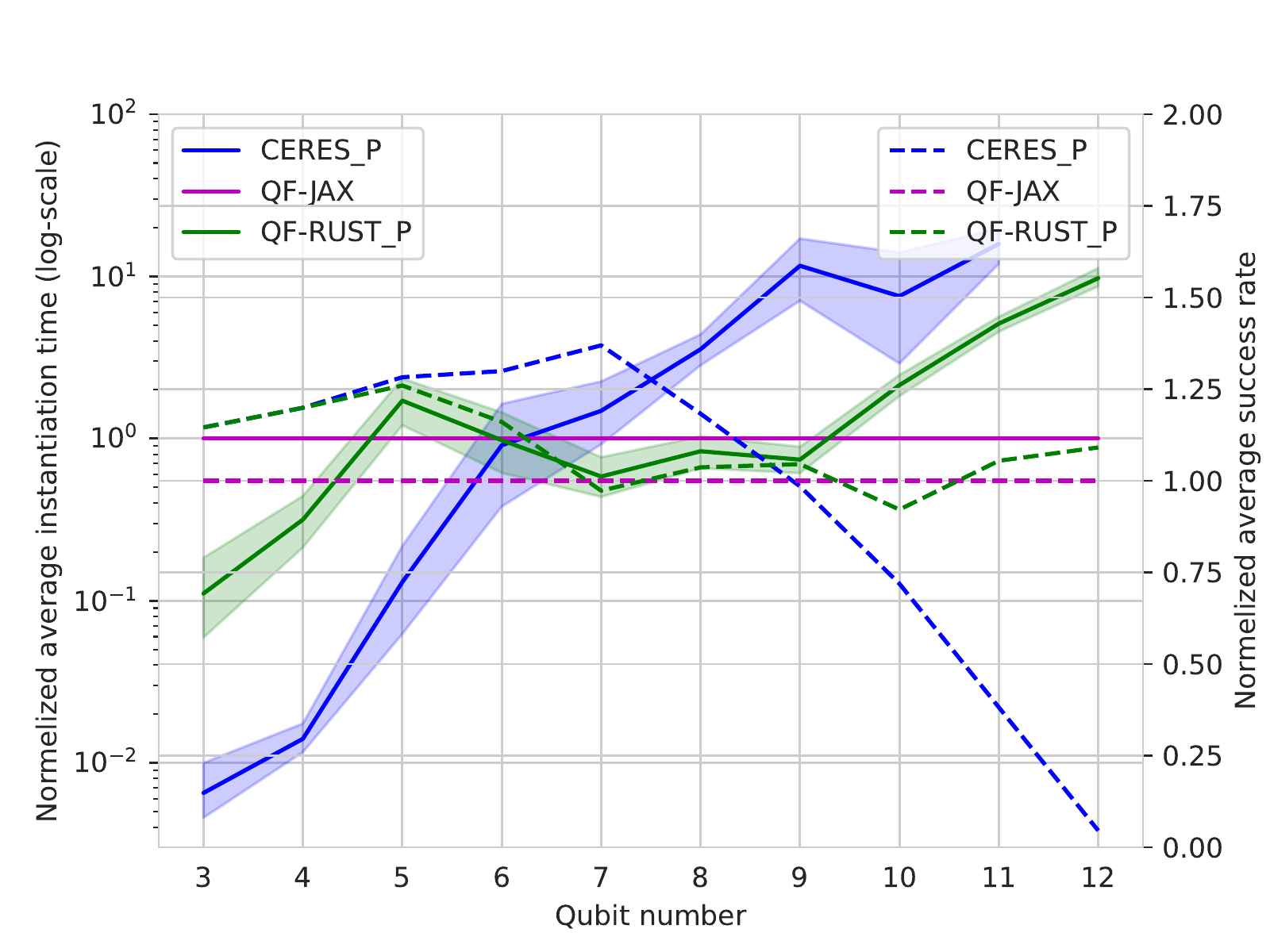}
    \caption{\footnotesize \it Average instantiation time normalized to QF-JAX instantiation time (left-hand side y-axis), together with normalized success rate (right-hand side y-axis), showing the strength of \qfacs for  larger circuits.}
    \label{fig:install}
    \end{minipage} &
    \begin{minipage}{\columnwidth}
    \includegraphics[width=\columnwidth]{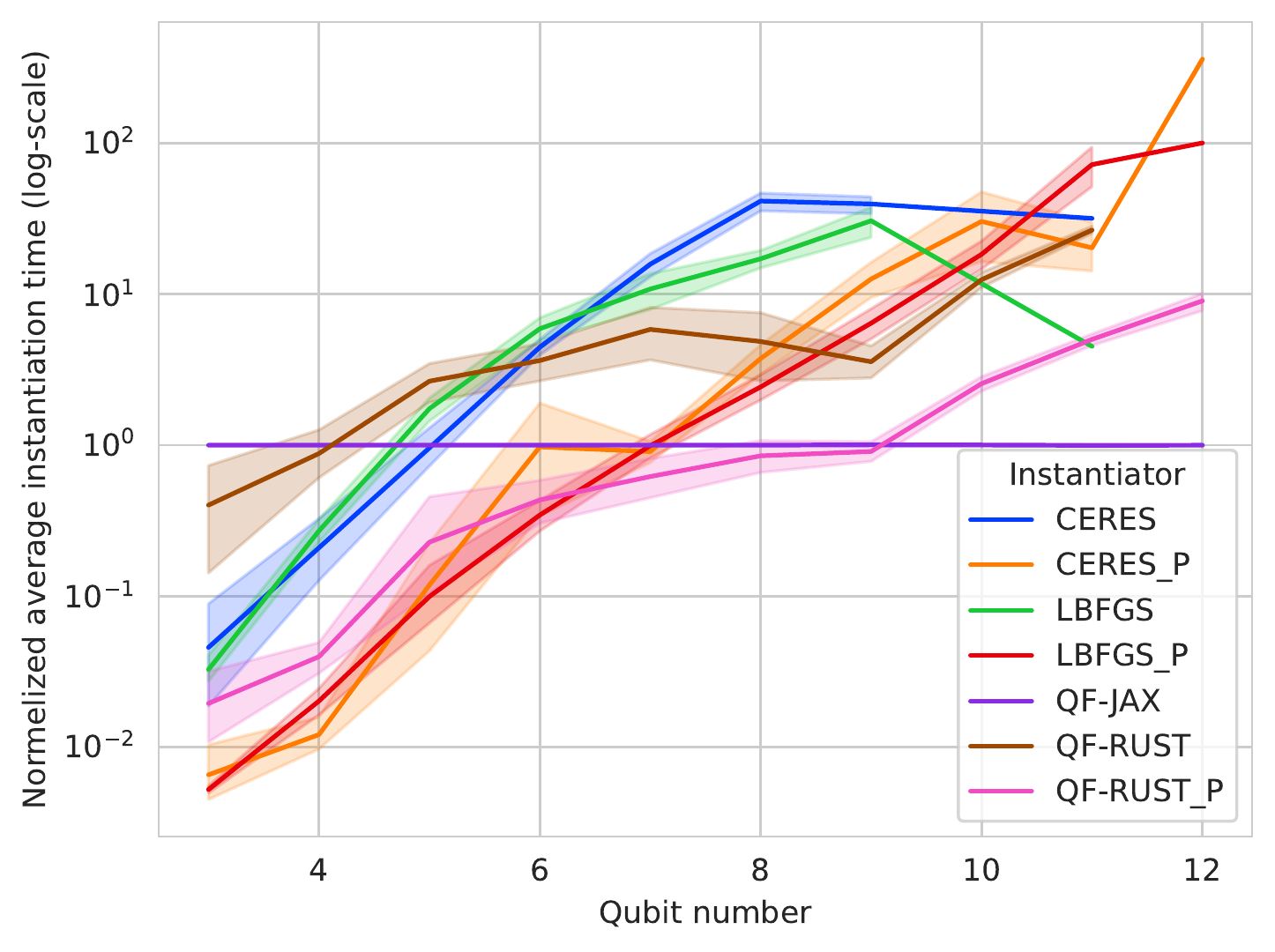}
    \caption{\footnotesize \it Average instantiation time normalized to QF-JAX instantiation time, in log scale. Lower is faster. While there are some differences in serial execution time between LBFGS and CERES, these wash out in parallel executions (LBFGS\_P and CERES\_P).}
    \label{fig:insttime}
\end{minipage} \\

\end{tabular}
\end{figure*}

We study \qfac's runtime and success rate efficacy by using it as a standalone instantiation tool on a set
of circuits generated by partitioning the input set in
Table~\ref{tab:benchmarks} into ``tiles'' ranging from three
to  12 qubits.
The dataset comprises of 1727 partitions obtained by randomly selecting 10 partitions
for every benchmark and partition size. Figures~\ref{fig:install}~and~\ref{fig:insttime} together with Table~\ref{tab:inst_succ_rate} summarize our findings.

\begin{table}[b!]
\footnotesize
\centering
\begin{tabular}{|c||ccc|}
\hline
\multirow{2}{6em}{Partition size} &\multicolumn{3}{c|}{Instantiation name} \\

   &CERES\_P &  QF-JAX &  QF-RUST\_P \\
   \hline
3           &     1.00 &         0.87 &            1.00 \\
4           &     1.00 &         0.83 &            1.00 \\
5           &     1.00 &         0.78 &            0.98 \\
6           &     0.97 &         0.74 &            0.86 \\
7           &     0.93 &         0.68 &            0.66 \\
8           &     0.69 &         0.58 &            0.60 \\
9           &     0.50 &         0.51 &            0.53 \\
10          &     0.30 &         0.42 &            0.38 \\
11          &     0.10 &         0.27 &            0.29 \\
12          &     0.01 &         0.14 &            0.16 \\
\hline
\end{tabular}
\caption{\footnotesize \it Instantiation success rate comparison. \qfacs has a better succuss rate in partitions with more than 8 qubits. For the small partitions, the platue detection mechanism in \qfacs is the main cause of unsuccessful intantiation.}
\label{tab:inst_succ_rate}
\end{table}

We give each optimizer a time budget of 10 minutes for circuits with up to 8 qubits and 2 hours for all the rest. We set the distance target $|| u_1 \cdot \ldots \cdot u_k - V ||$ to $10^{-10}$, as previous studies~\cite{qsearch,quest} have shown that this is
enough to guarantee good output fidelity when running on NISQ devices, as well as when simulating a perfect QPU. We measure the success rate of each optimizer to find a solution within the time budget.

Figure~\ref{fig:install} shows a comparison of the average
instantiation time plotted against the left-hand side {\it y-}axis and
success rate plotted against the right-hand side {\it y-}axis. The results indicate clearly that for the high volume circuits, in our data set, for partitions with more than {\it six} qubits \qfacs outperforms CERES in performance, and for partitions with more than {\it eight} qubits \qfacs also has a better success rate. At the bigger partition sizes, the GPU implementation of \qfacs outperforms the Rust based CPU implementation by an order of magnitude. A detailed comparison of the success rate is given in Table~\ref{tab:inst_succ_rate}.

For circuits with fewer than {\it six} qubits in our data set, the behavior is more nuanced. For the {\it three} qubit circuits, results
clearly indicate that CERES offers the best behavior. For the circuits
with {\it four} to {\it six} qubits, the results seem to indicate that
CERES is also best, e.g. 4-5 times faster and with a 2\%-22\% better success rate for five qubit circuits.

However, these results are skewed by the procedure used to generate
the benchmarking input set: we draw small ``circuits'' from partitions
of high qubit count circuits associated with proper algorithms,
e.g. tfim400. Due to the circuit structure of the algorithms, our
sample data set contains mostly circuits with a relatively low
volume. For example, the average number of U3s and CNOTs in our sample of
five-qubit circuits is 50 and 38 respectively. In contrast, the four-qubit hub4 Hubbard model circuit contains 155 and 180 U3s and CNOTs respectively.

For small volume circuits, CERES indeed does best in terms of success
rate and performance. For high volume circuits \qfacs offers much
better performance, again by orders of magnitude.  For example for the
{\tt hub4} Hubbard model circuit, which uses {\it four} qubits and
contains 335 gates, \qfacs is 5 times faster. The success
rates of \qfacs and CERES are comparable across all four and five
qubit circuits we studied, with slight differences for six qubit
circuits. Most of the differences can be attributed to the plateau
detection heuristic in \qfac, which is not yet well tuned. As using
numerical optimizers requires judicious tuning of their parameters, we
expect to be able to improve \qfacs convergence.

We also note that for small-volume partitions, the GPU implementation of \qfacs has significant overhead due to the JIT compilation of JAX, and booting up the GPU.

Interestingly, because the GPU implementation continues to iterate
over all the multistarts in parallel, even when in some of them the plateau
detection heuristic is triggered, the GPU might be able to escape local
minima, where it wasn't an option in the sequential flow. We actually
see this happening in our experiments, where the GPU implementation
improves the quality of results over the Rust implementation although
inherently they do the same exact calculations.

Overall, the data indicates that when doing instantiation using
parameterized gate encodings, CERES and LBFGS are the preferred
solution for any three-qubit circuit. For four- to six-qubit circuits,
the volume matters.  For low volume, CERES and LBFGS are
preferred. For mid-volume  circuits, if speed is a concern, the CPU
based implementation of \qfacs is recommended, while there are still
unexplored trade-offs with respect to the quality of the solution. The
GPU implementation of \qfacs is the only solution for high volume
circuits.

\subsection{Circuit Optimization Scaling}\label{sec:synth_eval}

We have integrated \qfacs into the re-synthesis gate deletion workflow
presented in~\cite{qce22_ed}. The flow first partitions the circuit,
and then performs a uni-directional sweep trying to delete one gate at
a time while re-instantiating the reduced partition to its original
unitary. This divide-and-conquer technique transforms compilation into
an embarrassingly parallel task.  In this context, we are interested
 to understand  the scaling of large circuit
compilations, as well as assessing \qfac's efficacy as a numerical
optimizer, which is captured directly by the number of deleted gates
within a circuit.

We use four big circuits (vqe12, adder63, heisenberg64 and shor26) and
execute on up to six Perlmutter CPU nodes and on one to 36 GPUs for the
GPU\footnote{36 GPUs account for nine hybrid nodes on Perlmutter. CPU
nodes are different from hybrid nodes, but these differences are
unlikely to affect results as the jobs are embarrassingly parallel.}
implementation. As \qfacs can handle high volume circuits, this allows
us to partition the input circuits into panels of increasing
volume, ranging from three to nine qubits, and the number of gates in the partitions ranges up to 515, similar in size to some of the original benchmark circuits, see Table~\ref{tab:benchmarks}. The average partition gate count is $66$ for partitions of 7 qubits. In these experiments, we set a time limit of 11 hours for the compilation jobs.

\begin{figure}[ht]
    \centering
    \begin{subfigure}{0.24\textwidth}
        \includegraphics[width=\textwidth]{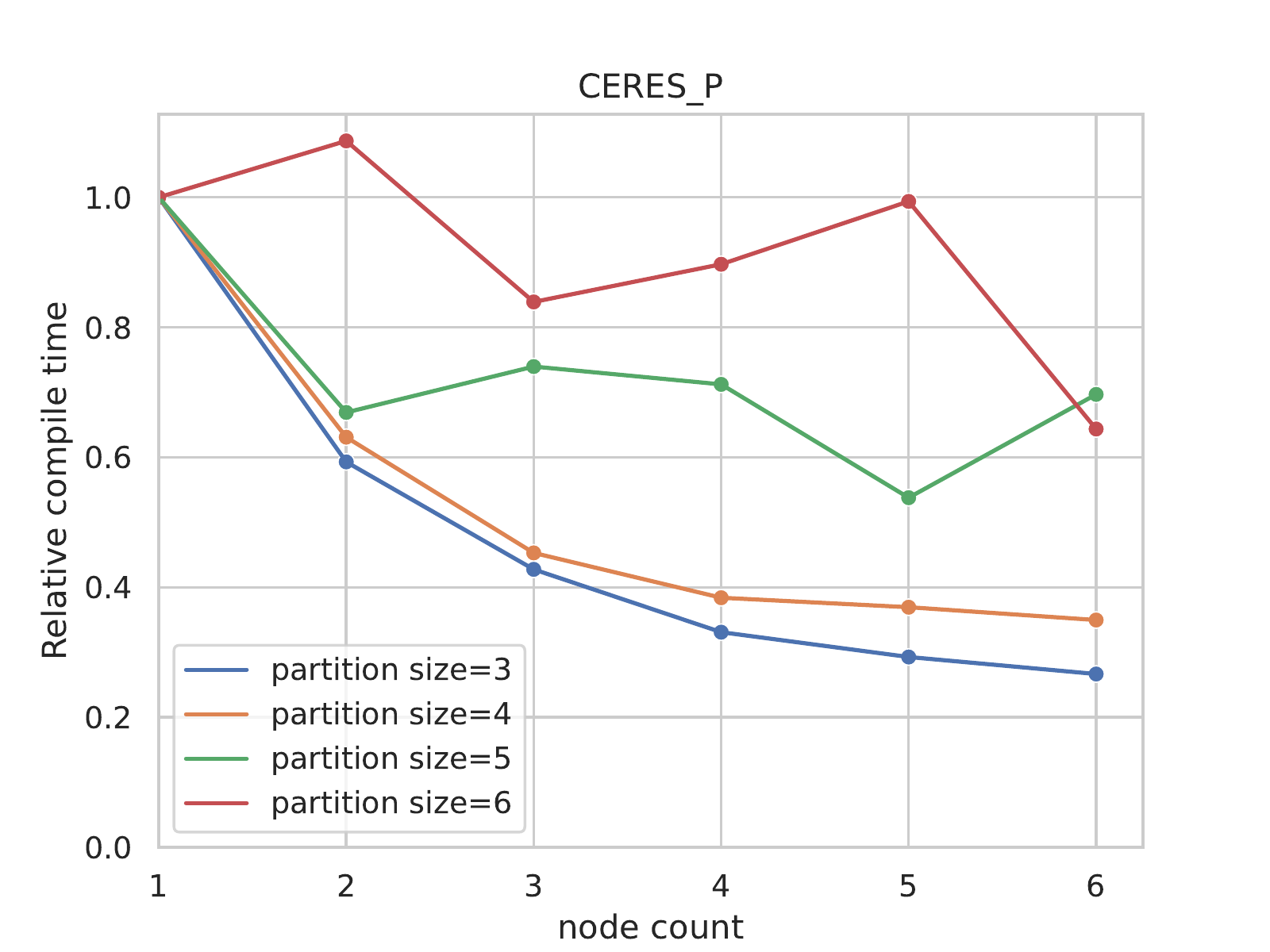}
    \end{subfigure}
    \begin{subfigure}{0.24\textwidth}
        \includegraphics[width=\textwidth]{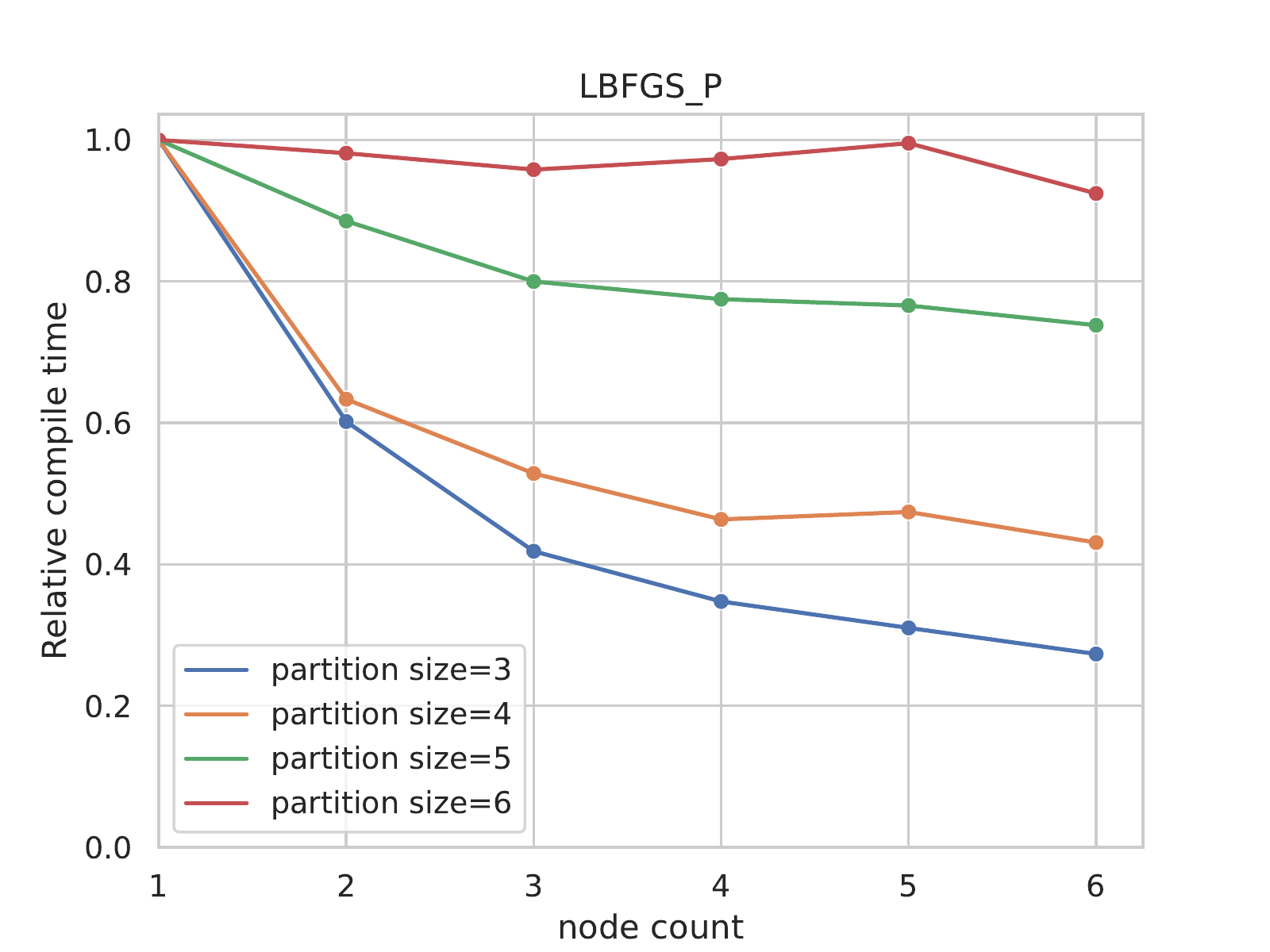}
    \end{subfigure}
    \begin{subfigure}{0.24\textwidth}
        \includegraphics[width=\textwidth]{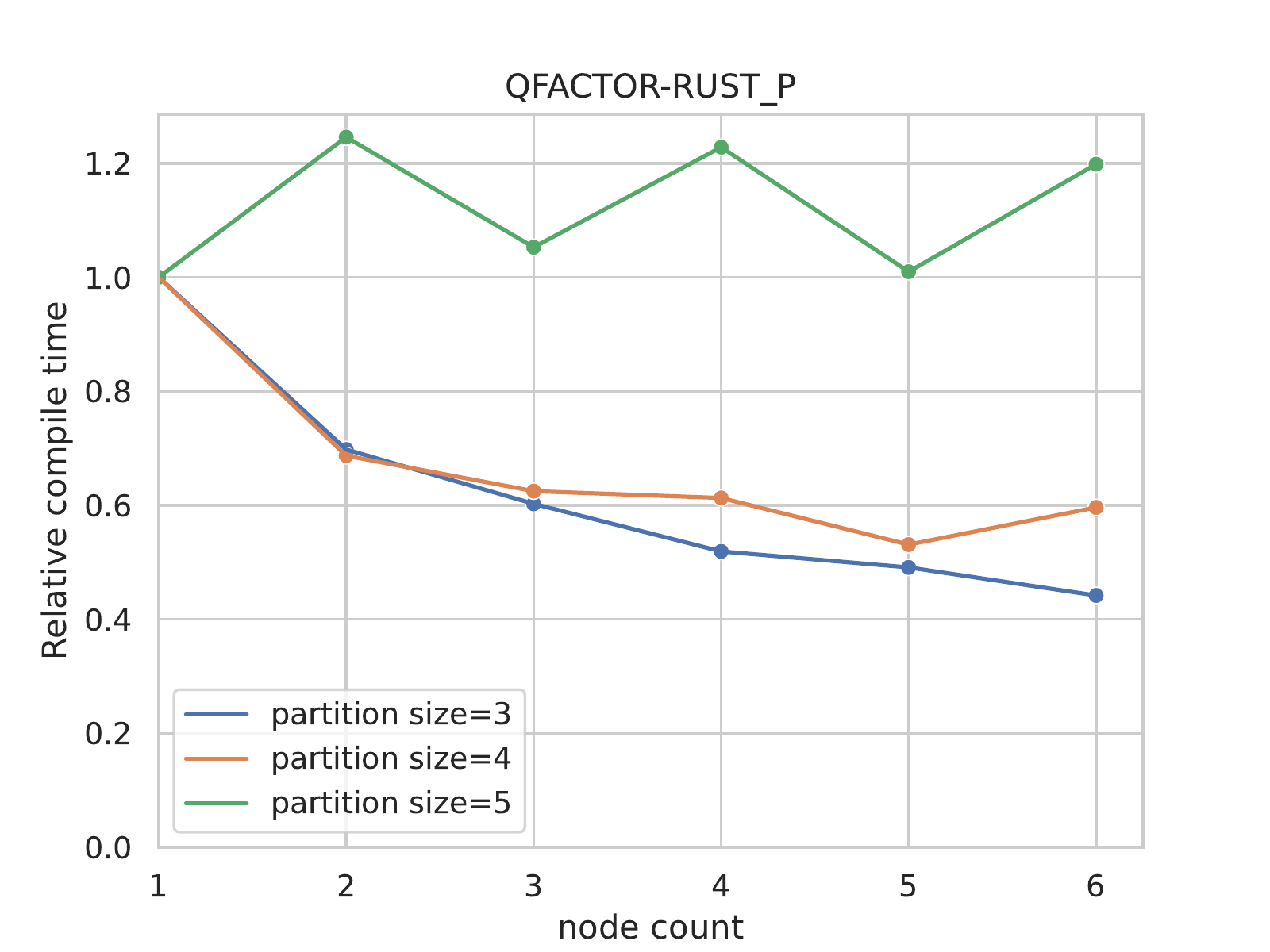}
    \end{subfigure}
    \begin{subfigure}{0.24\textwidth}
        \includegraphics[width=\textwidth]{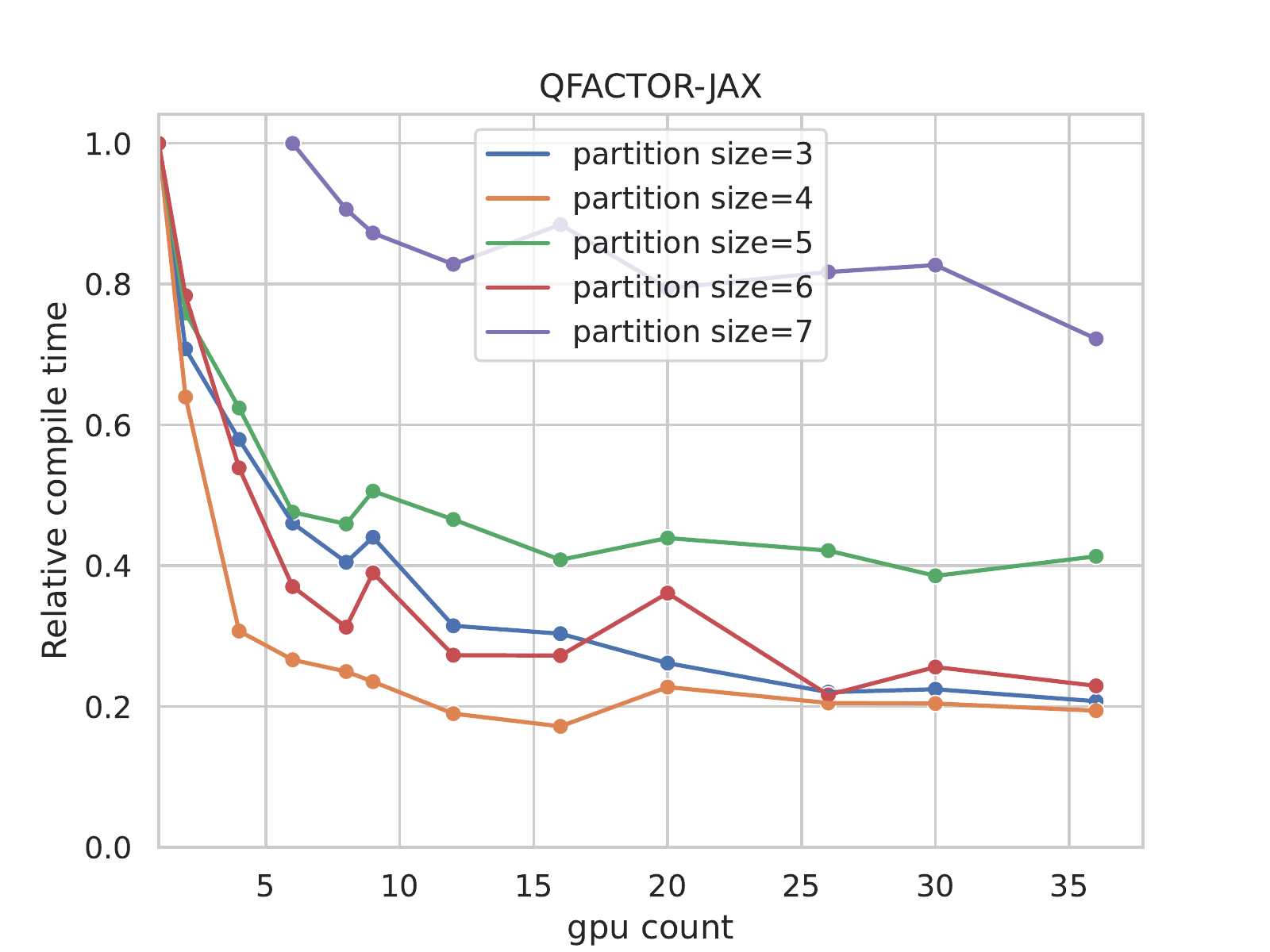}
    \end{subfigure}
    \caption{\footnotesize \it Strong scaling of compilation of large circuits. Relative compilation run time of different instantiators, across various partition sizes with varying resources, showing strong parallelization. Performance is normalized to either 1 CPU node, or to 1 GPU device respectively, unless  it timed out, then it's for 2 resources.}
    \label{fig:strong_par}
\end{figure}

Figure~\ref{fig:strong_par} shows the performance scaling
results. Most of the trends are explained by the instantiation
performance results,  for
brevity we omit a detailed evaluation. Overall, for the circuit
volumes they work best, all optimizers enable  strong scaling of the compilation workflow. The
strong scaling limit is determined by the number of partitions in the
large circuit that is compiled: jobs that partition large circuits into small panels
scale ``better'' than jobs that use larger partitions. Strong
scaling for CPU based instantiators seems to be sensitive to load
balancing between the workers. This sensitivity does not appear in the low volume partitions regime where they are at their best. The GPU based implementation shows the best scaling and it seems insensitive to load balancing issues.

\subsection{Circuit Optimization  Efficacy}\label{sec:synth_qual}

 To assess
efficacy we measure the number of gates deleted by this pass for
increasing partition size and report their average.  For each flow configuration, we use the same input set of circuits. Due to the 11
hours compilation time limit and the differences in execution time
between optimizers, we note that increasing the partition size
decreases the number of successful jobs due to timeouts. We also don't initiate a circuit compilation  with a bigger partition size, if a lower one timed out. Furthermore,
for a given partition size we note that a different set of circuits
may be successfully optimized by a flow configuration.

\begin{figure}[htb]
    \begin{subfigure}{0.24\textwidth}
        \includegraphics[width=\textwidth]{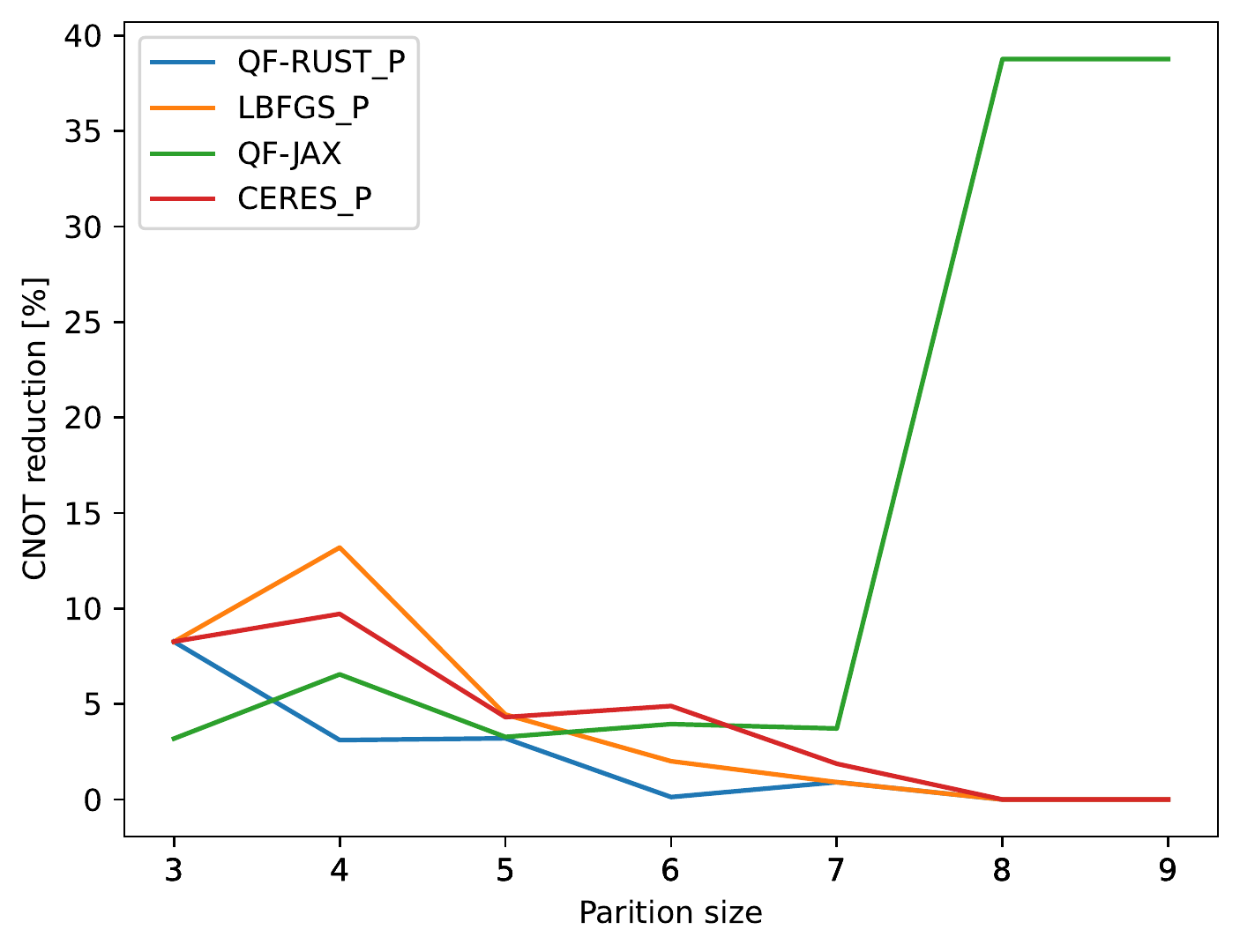}
    \end{subfigure}
    \begin{subfigure}{0.24\textwidth}
        \includegraphics[width=\textwidth]{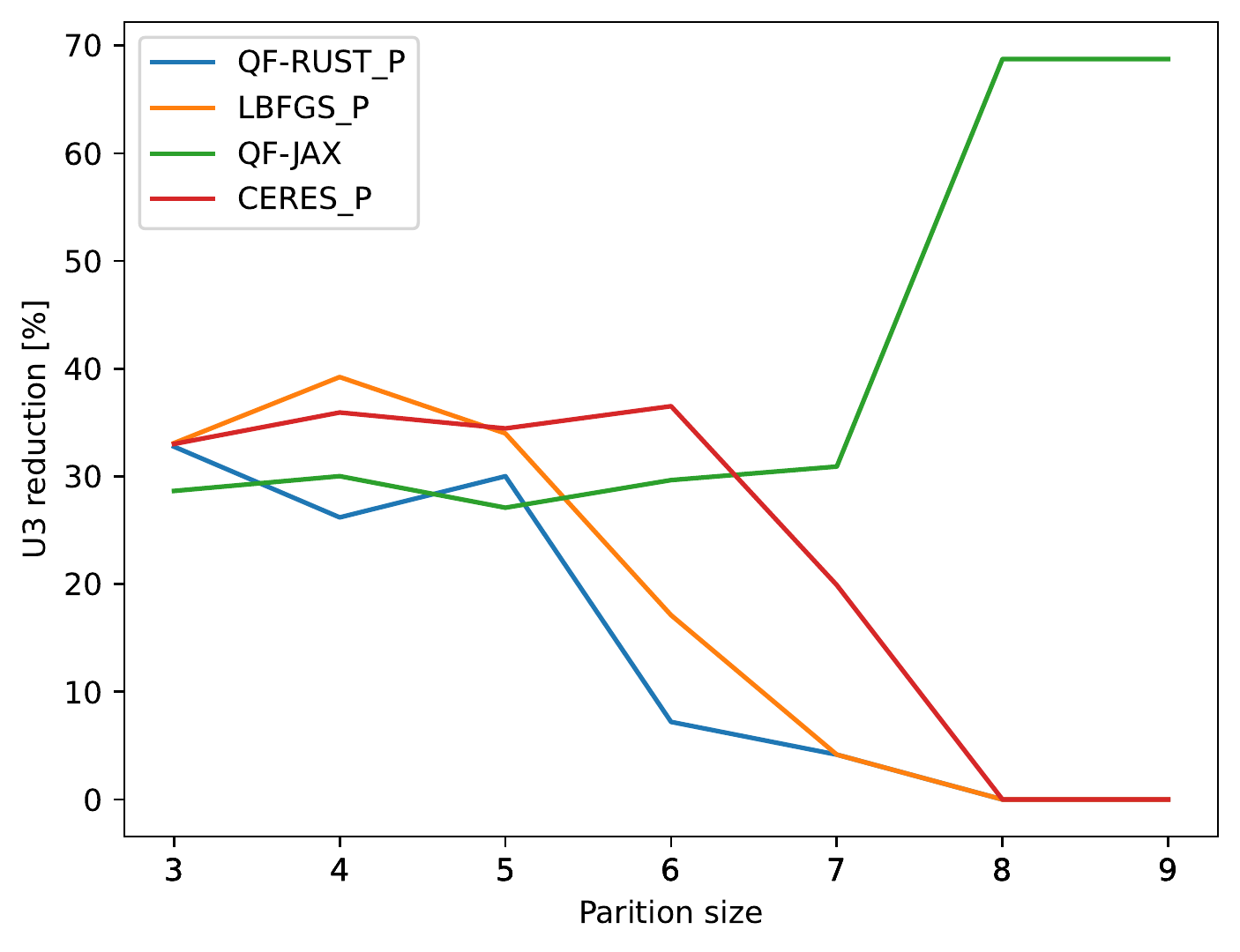}
    \end{subfigure}
    \caption{\footnotesize \it Average percentage of CNOT and U3 reduction in the gate deletion flow.}
    \label{fig:reduction_del_flow}
\end{figure}

Figure~\ref{fig:reduction_del_flow} presents the average reduction in
U3 and CNOT gates when increasing the number of qubits within a
partition. As shown, the ability to process nine qubit partitions
leads to $4\times$ and $2\times$ increases in the number of U3 and
CNOT gates deleted, respectively. The GPU implementation of \qfacs is the only instantiator that didn't cause a time out for partition sizes of 8 and 9.  For three- to five-qubit partitions
the quality of output is comparable. We also note that \qfacs and
LBFGS particularly struggle on the Hubbard model circuits, while CERES
is able to find a more efficient implementation. For \qfacs 
hyperparameter tuning may be used to  improve efficacy.

\begin{figure*}[htbp!]
    \begin{subfigure}{0.5\textwidth}
        \includegraphics[width=\textwidth]{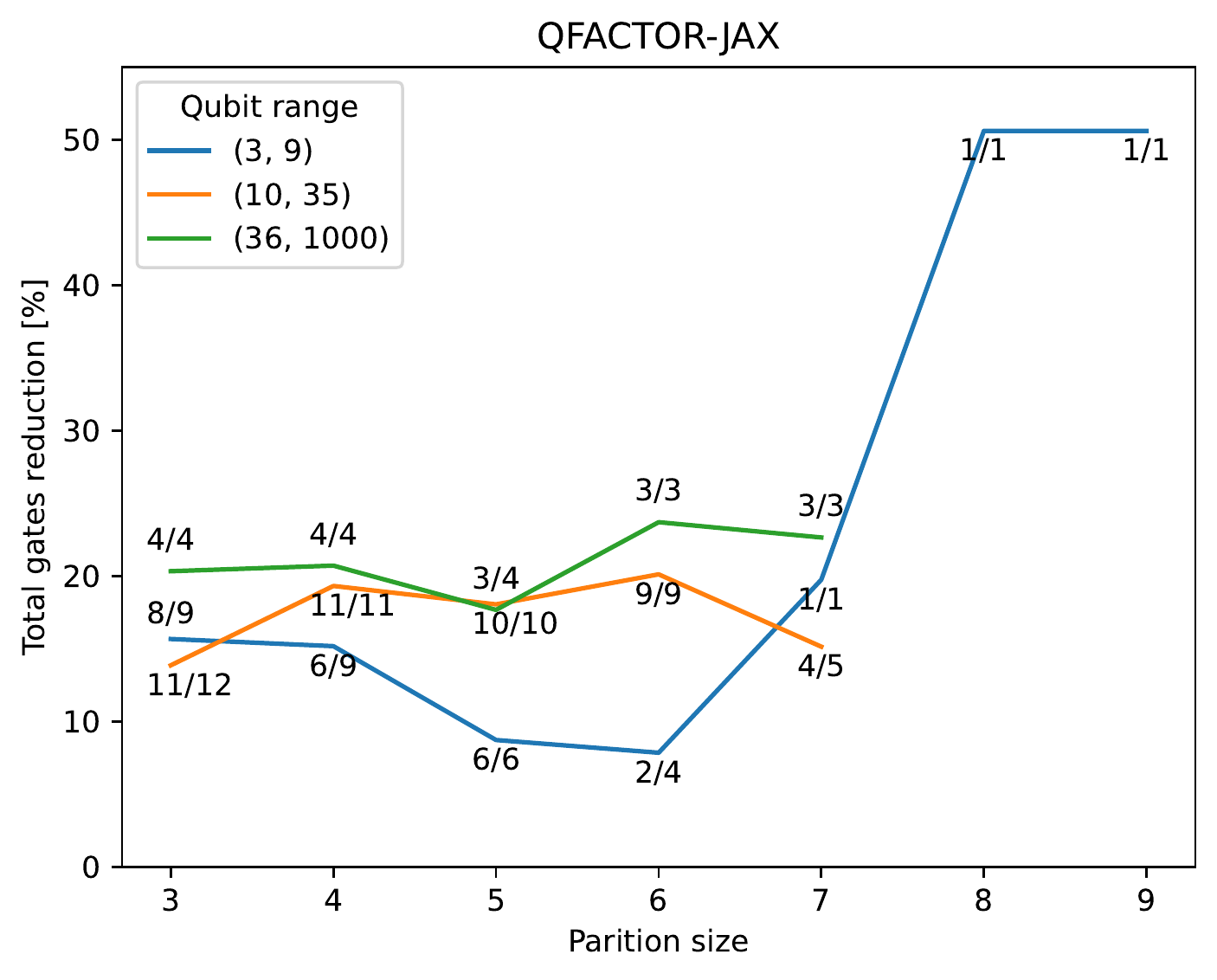}
    \end{subfigure}
    \begin{subfigure}{0.5\textwidth}
        \includegraphics[width=\textwidth]{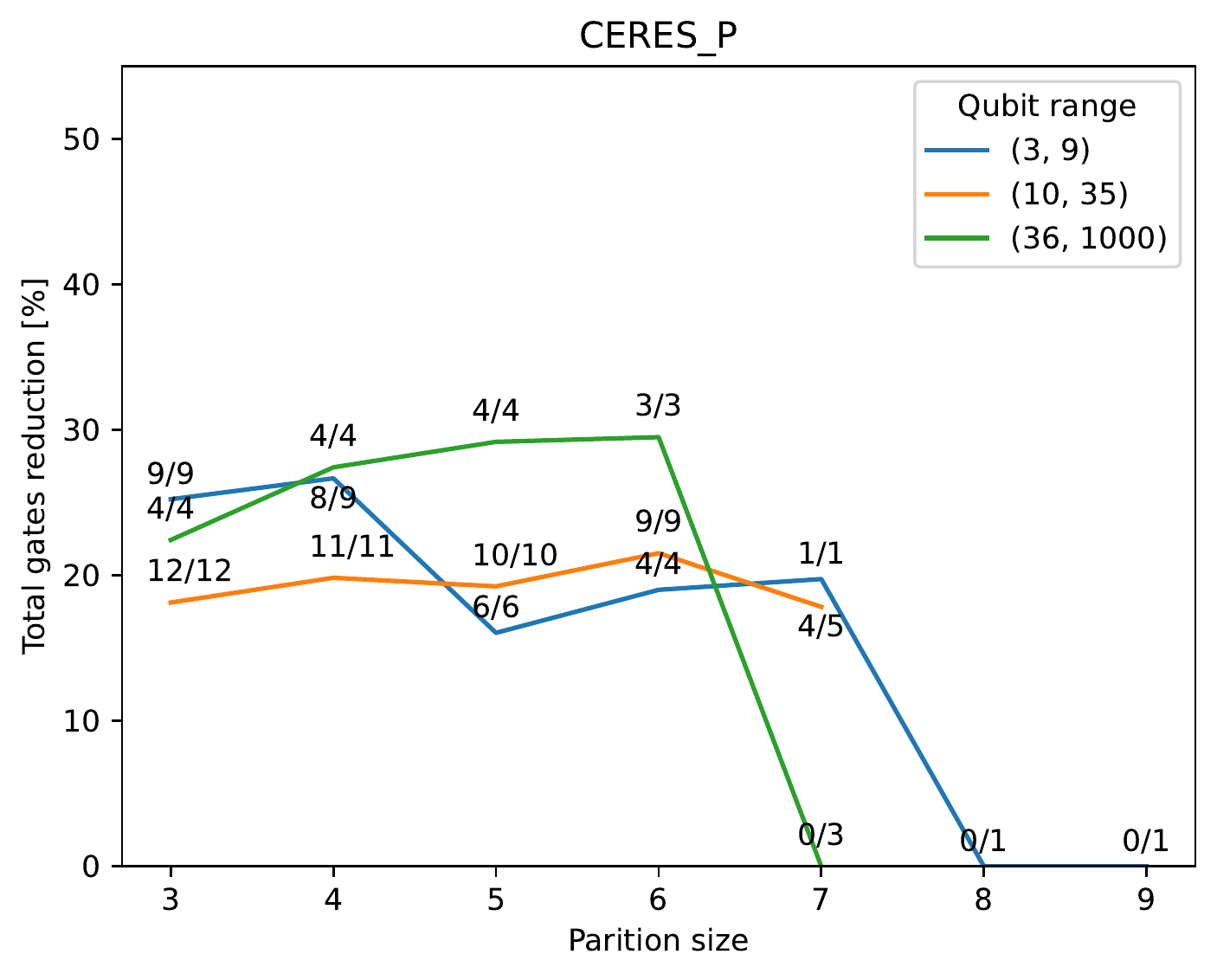}
    \end{subfigure}
    \caption{\footnotesize \it Average percentage of total gate reduction in the gate deletion flow divided into buckets according to the amount of qubits in the circuit, CERES\_P and QF-JAX are shown. For each data partition size and bucket, also show how many circuits were successful.}
    \label{fig:buckets}
\end{figure*}

For a more detailed understanding of the dynamics, we split the input
circuits by size into three buckets: small - up to 9 qubits, medium -
10-36 qubits, and large - 36 and above qubits. The data is presented
in Figure~\ref{fig:buckets}.  For the small circuits, due to its
better scalability \qfacs reduces gate count by an ``average'' of 40\%,
and CERES by an ''average'' of 25\%.

The dynamics for 3- to 6-qubit partitions for both optimizers are
explained by the composition of the data set: for small circuits a
partition may cover a very large fraction of the original circuit,
thus capturing complex physical interactions present in the domain science problem formulation. On the other hand, when partitioning a larger
circuit, the partitions are likely to be themselves "simple"
circuits. Intuitively, the bigger the fraction of a circuit a
partition covers, the bigger the chance it may capture complex
interactions between qubits as determined by the domain science behind
the particular algorithm.  This also explains some of the differences
in the quality of optimization between the medium and large circuits
buckets. The impact of circuit size and the time limit can also be
observed: large circuits have more partitions and their jobs time out
more when increasing partition size, as they form more complex
partitions.

\begin{table}[hb]
\tiny
\centering
\begin{tabular}{ |c| c| c| }
\hline
 Circuit & U3  & CNOT \\ 
 \hline

{\tt adder9 }& 64 & 98 \\
{\tt add17}& 348&232\\

{\tt adder63 }& 2885 & 1405 \\
\hline
{\tt mult8}& 210&188\\
{\tt mult16}& 1264&1128\\
\hline
{\tt grover5 }& 80 & 48 \\
{\tt hhl8}& 3288 &2421\\
{\tt shor26 }& 20896 & 21072 \\
\hline
{\tt hub4}& 155 & 180 \\
{\tt hub18 }& 1992 & 3541 \\
\hline
{\tt heis7}& 490&360\\
{\tt heis8}& 570&420\\
{\tt heis64}& 5050&3780\\
\hline
{\tt tfim8}& 428&280\\
{\tt tfim16}& 916&600\\
{\tt tfim400 }& 88235 & 87670\\
\hline
{\tt qae11}& 176&110\\
{\tt qae13}& 247&156\\
{\tt qae33 }& 1617&1056\\
{\tt qae81}& 7341 & 4840\\
\hline
{\tt qaoa5 }& 27 & 42 \\
{\tt qaoa10 }& 40 & 85 \\
{\tt qaoa12 }& 90 & 198 \\
\hline
{\tt vqe5}& 132&91\\
{\tt vqe12}& 4157&7640\\
{\tt vqe14}& 10792&20392\\
\hline
{\tt qpe8}& 519&372\\
{\tt qpe10}& 1681&1260\\
{\tt qpe12}& 3582&2550\\

\hline
\end{tabular}
\caption{\footnotesize \it Benchmarks and their gate counts, upper bound of $\approx 200,000$.  The name suffix represents the number of qubits in the circuit, up to 400 qubits. }
\label{tab:benchmarks}
\end{table}

\section{Discussion}
\label{sec:discussion}

The results indicate that \qfacs greatly accelerates the performance
of instantiation and it enables us to handle larger volume circuits.
Both general-purpose optimizers and \qfacs scale exponentially with
parameters. Instantiation with GPO algorithms use parameterized gates:
CERES and LBFGS formulations treat an U3 gate as three parameters. \qfacs treats
it as a single parameter benefiting from memory reduction and is most
likely simpler to optimize objective functions.  Moreover, \qfacs can
be generalized to treat multi-qubit ($n > 2$) unitaries as a single
parameter for further scalability improvements, while GPOs still have
to represent these with $O(2^n)$ parameters.
This generalization is likely to improve the performance of \qfacs even further.

Figure~\ref{fig:reduction_del_flow} shows the evolution with
partition size of the quality of solutions generated by the gate
deletion workflow.  These clearly indicate that when incorporated into
circuit optimization compiler passes, \qfacs improves the
quality of the generated circuit when compared against general-purpose optimizers.

The observed dynamics of compilation time
and output circuit quality with \qfacs enable us to draw very useful
conclusions with respect to architecting instantiation or synthesis
based workflows for large circuits.

The improvement in gate count reduction seems to grow with the
partition size, while numerical optimization algorithms usually scale
exponentially.  This opens the door for configuring compiler workflows
based on tradeoffs between output quality and compilation
speed. Furthermore, if compilation speed is important, generating
three qubit partitions seems to provide a good default.

Another very interesting question is finding the partition size that
maximizes the quality of large circuit optimizations: {\it Is it a
function of circuit size and structure?}  The answer may guide future
research attempts to scale  direct unitary instantiation (numerical
optimization) and synthesis algorithms (search over circuit structures
and numerical optimization) with qubits.

A partial answer provided by this study is that, for circuits with
many qubits, the efficacy of circuit optimizations using instantiation
saturates with partitions of 5-7 qubits. These partitions seem to be
the sweet spot for numerical optimization with \qfac or CERES.  While
for this submission we show results for gate deletion in large
circuits using partitions up to 9 qubits, we have started experiments
for partitions up to 12 qubits to understand better the \qfacs
dynamics.

This needs more investigation, and the reasoning is more subtle. The
real metric is probably the ratio of circuit gates to partition gates
rather than number of qubits, augmented by structure imposed by domain
science. Algorithms are written in terms of qubit
interactions. Whenever a partition captures the totality of
interactions between groups of qubits, direct optimization will see
great benefits. Otherwise, we expect to see the same saturation in the
quality of the circuit optimization.

Based on these considerations, and to conserve the limited Perlmutter
supercomputing allocation, we have stopped our instantiation experimentation at
partitions with only 12 qubits. The experimental data clearly
indicates that \qfacs is able to process even larger circuits. Note
that we have used +20K hybrid GPU node hours and +13K CPU node
hours. This amounts to $20K\times(4GPUs + 64\ cores)$ and
$13K\times128\ cores$ physical resources. We thank the NERSC directorate for
allowing us to run on their discretionary allocation.

Due to the tensor network formulation, our algorithm formulation
benefits greatly from GPU acceleration.  We have not tuned at all the
GPU resource allocation strategy for compilations of four- to 12 qubit
circuits. As both the number of qubits and the number of gates
determines performance, additional improvements are likely to be had
by a more judicious resource allocation on GPUs.

Numerical optimization behavior is notoriously fickle and requires
customization for the problem at hand. For this study we performed
only minimal tuning of the \qfacs parameters. As each instantiated
circuit has a different structure, therefore leading to a different
optimization surface, we believe we can improve both performance and
quality of optimization with parameter tuning.

The results were demonstrated on a U3+CNOT gate set. \qfacs is
easily portable to other gate sets, all that it requires is providing
derivatives.  An interesting question remains if adopting a different
gate set changes the conclusions of this paper.

\section{Conclusion} \label{sec:conclusion}

We introduce \qfac, a domain specific optimizer designed for quantum circuit instantiation problems. When compared against general purpose optimizers \qfacs scales  better and is able to process much larger circuits.  When replacing general purpose optimizers with \qfacs into
a compilation workflow, this capability translates into better optimized circuits.  Due to its formulation \qfacs is able to execute on GPUs and benefit from their hardware acceleration. Furthermore, the \qfacs enabled compiler running on GPUs seems to scale out linearly in distributed memory environments; this does not happen for CPU based compiler workflows, regardless of the numerical optimizer. This bodes well for the more general adoption of synthesis based quantum circuit optimizations: while their efficacy in  optimizing circuits is recognized, adoption is hampered by runtime overhead.  By significantly improving circuit quality and scaling in distributed memory GPU based environments \qfacs provides a first good step into alleviating execution overhead concerns.

\section*{Acknowledgements}

The research presented in this paper (LC) was supported by the Laboratory Directed Research and Development (LDRD) program of Los Alamos National Laboratory (LANL) under project number 20230049DR.  CI was supported by the U.S. DOE under contract DE5AC02-05CH11231, through the Office of Advanced Scientific Computing Research (ASCR), under the Accelerated Research in Quantum Computing (ARQC) program.

This research used resources of the National Energy Research
Scientific Computing Center (NERSC), a U.S. Department of Energy
Office of Science User Facility located at Lawrence Berkeley National
Laboratory, operated under Contract No. DE-AC02-05CH11231 using NERSC
award DDR-ERCAPm4141.

\bibliographystyle{ieeetr}
\bibliography{quantum,bibliography}
\clearpage

\end{document}